\documentclass[aps,prb,twocolumn,showpacs,preprintnumbers,amsmath,amssymb,superscriptaddress,floatfix]{revtex4}

\usepackage{graphicx}
\usepackage{dcolumn}
\usepackage{bm}
\usepackage{stmaryrd}
\usepackage{latexsym}
\usepackage{amssymb}
\usepackage{amsfonts}
\usepackage{amsmath}
\usepackage{fancybox}
\usepackage{color}

\definecolor{MyDarkBlue}{rgb}{0,0.08,0.45}

\begin{document}
\title{Stoner ferromagnetic phase of a graphene in the presence of an in-plane magnetic field}
\date{\today}
\author{A. Qaiumzadeh}
\affiliation{School of Physics, Institute for Research in Fundamental
Sciences, (IPM) 19395-5531 Tehran, Iran}
\affiliation{Institute for Advanced Studies in Basic Sciences
(IASBS), Zanjan, 45195-1159, Iran}
\author{R. Asgari}
\email{ asgari@theory.ipm.ac.ir}
\affiliation{School of Physics, Institute for Research in Fundamental
Sciences, (IPM) 19395-5531 Tehran, Iran}
\begin{abstract}
We study the effects of an in-plane magnetic field on
the ground state properties of both gapless and gapped graphene sheets
within Random Phase Approximation. The critical magnetic field which leads to a fully spin polarized phase increases by decreasing the carrier density at zero gap indicating that no spontaneous magnetic phase transition occurs. However, at large energy gap values it decreases by decreasing the density. We find a continuous quantum magnetic phase transition (Stoner phase) for Dirac fermions in a doped
graphene sheet. Novel in-plane magnetic field dependence of the charge and spin susceptibilities are uncovered.
\end{abstract}

\maketitle

\section{Introduction}

Graphene is an atomically thin two-dimensional (2D) electron system composed of carbon atoms
on a honeycomb lattice. Several experimental groups have recently~\cite{geim} introduced techniques which enabled isolation and
study of systems with one or a small number of graphene layers.
The interesting physics of graphene systems stems from the fact that its envelope-function of low-energy
Schr\"odinger equation is equivalent to the massless 2D
Dirac equation.  In the case of graphene the spinor structure in the Dirac equation
refers to sublattices of its honeycomb structure and its Brillouin-zone valleys, instead of the spin degrees of freedom and the
electron-positron picture. Graphene, therefore, presents a new type of
many-body problem in which the noninteracting low energy quasiparticle dynamic~\cite{review}
is effectively described by a 2D massless Dirac Hamiltonian $\hat{H}_0=\hbar
v_{\rm F}{\overrightarrow{\sigma}}\cdot{\bf k}$, with two chiral eigenvalues, $\pm \varepsilon_{\bf
k}$ where $\varepsilon_{\bf
k}=\hbar v_{\rm F}|{\bf k}|$, with
$v_{\rm F}\simeq10^6$m/sec, is the Fermi velocity of carriers.
The chirality plays an important role in the novel electronic properties of graphene.

The strength of interaction effects in an ordinary two-dimensional electron gas ( 2DEG) increases with decreasing the carrier density. At low densities, the effective velocity is suppressed, the charge compressibility changes sign from positive to negative, and the spin-susceptibility
is strongly enhanced.  In the Dirac-like electrons in graphene, it has been
shown~\cite{asgari,polini} that the velocity is enhanced rather than suppressed, and that
the compressibility, always remains positive, and the
spin-susceptibility is suppressed. These
qualitative differences are due to the exchange interactions between electrons near the
Fermi surface and electrons in the negative energy sea.
The interband excitations are closely
analogous to the virtual particle-antiparticle excitations of a truly relativistic electron gas.

Conventional 2DEG has been a
fertile source of surprising new physics for more than four decades.
In recent years, because of the important
and novel physical properties found in both theoretical and
technological applications, there has been a large amount of theoretical and
experimental studies on the effects of parallel magnetic field $B$
in a 2DEG. A great deal of activity was spawned in the last decade to
understand the apparent metal-insulator transition observed in
Si-MOSFET and GaAs based structures.\cite{abrahams} Although
the basic mechanism and the existence
of a quantum phase transition is still a matter of on-going debate,
experiments have amassed a wealth of data on the transport
properties of the 2D electron systems in the metallic state.
Zhang and Das Sarma\cite{sarma1} investigated the
ground-state properties of the 2DEG in the presence of an in-plane
magnetic field $B$ using random phase approximation (RPA).
They showed that for small Wigner-Seitz density parameter $r_s=(\pi n a^2_B)^{-1/2}$ in
which $a_B$ is the Bohr radius in
the medium of interest and in the absence of the
magnetic field the system prefers to be in a paramagnetic state. As $B$
increases the ground-state energy is minimized at a special nonzero spin polarization
denoted by $\zeta^*$. The degree of spin polarization is defined as $\zeta=|n_\uparrow-n_\downarrow|/(n_\uparrow+n_\downarrow)$, and
$n_{\uparrow(\downarrow)}$ is electron density with spin up (down). When $B$ increases to a critical value $B_c$ in which the system is fully spin polarized, there exist two $\zeta^*$ values, smaller and equal to one, where the total
energy is minimized. They have shown that the first order phase transition from paramagnetic-to-ferromagnetic takes place. Importantly, beyond the critical field the energy minimum is at $\zeta^*=1$ and the system is fully spin polarized.

Suba{\c s}{\i} and Tanatar\cite{tanatar}, on the other hand, studied the
same system by using a parameterized expression
for the correlation energy provided by the quantum Monte Carlo (QMC)
simulations.\cite{attaccalite} They found that the 2DEG in the presence
of an in-plane magnetic filed $B$ ($B\geq B_c$) undergoes a first order phase transition
to the ferromagnetic state (Bloch ferromagnetism) in the density regions associated with
$0<r_s<7$ and $20<r_s<25$, while for $7<r_s<20$ their results
predicted a continuous phase transition (Stoner ferromagnetism).

In an electron gas system the physical observable quantities most directly related to the energy are the
compressibility which measures the stiffness of the system against changes in the density of electrons and the spin-susceptibilities.
In bulk electronic systems, the spin-susceptibility can usually be extracted successfully from
total magnetic susceptibility measurements. This is, however, likely to be challenging in the case of single-layer graphene. In the 2DEG, on the other hand, information about the spin-susceptibility can often be extracted from weak-field magneto-transport experiments using a tilted magnetic field to distinguish spin and orbital response.

Recently, Hwang and Das Sarma\cite{g_sarma} have shown that the in-plane magnetic field induces graphene magneto-resistance which is negative for intrinsic gapless graphene while for extrinsic gapless
graphene, magneto-resistance is a positive value at fields lower than
the critical magnetic field and negative above the critical magnetic
field. The effect of in-plane magnetic field on microwave magneto-transport~\cite{mani} in doped graphene is an open problem.

The purpose of this paper is to study the effects of in-plane magnetic field on disorder free doped graphene at zero temperature. These effects are very important and have some novel and unusual
properties in comparison to the conventional 2DEG.
We have revisited the problem of the compressibility and spin-susceptibilities in the presence of an in-plane magnetic field and find that the charge compressibility exhibits a crossover between paramagnetic and ferromagnetic phases depending on the carrier density and the gap values.
Before describing the details of the theory and presenting our results, we point out that the most novel electronic properties of graphene which will be discussed here are based on interband interaction and exchange interaction between electrons near the Fermi surface in graphene sheets.
\begin{figure}
\includegraphics[width=7.5cm]{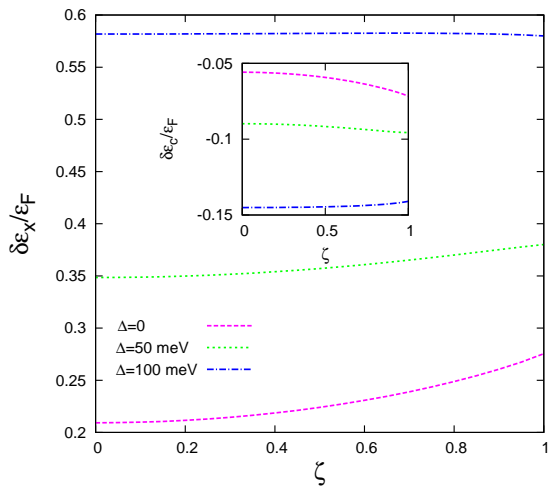}
\caption{(color online). Exchange energy as a function of degree of spin polarization, $\zeta$
for various gap energies. In the inset: the correlation energy as a function of $\zeta$ for various gap energies.} \label{}
\end{figure}

\begin{figure}
\includegraphics[width=7.5cm]{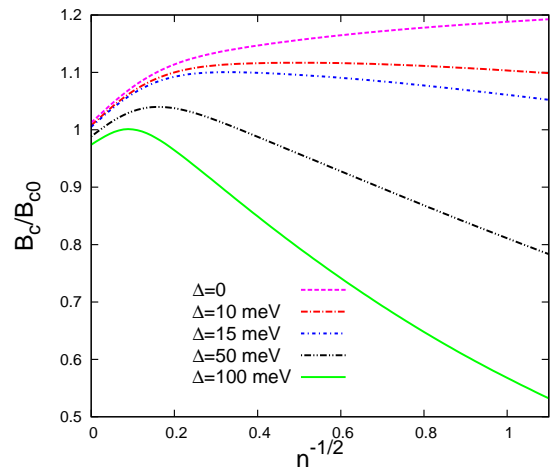}
\caption{(color online). Critical magnetic field as a function of inverse
square root of density ( in units of $10^{-6}$ cm) for various gap energies.} \label{}
\end{figure}

The content of the paper is as follows. In Section II we
discuss about our theoretical model which contains the Zeeman energy. Our numerical results are given in Section III. Eventually, Section IV contains the summery and conclusions.

\section{Theoretical model}

We consider a doped graphene sheet with a peculiar gap opening due to sublattice symmetry breaking where 2D massive
Dirac fermions at low energy is described by
noninteracting Hamiltonian~\cite{qaium} $\hat{H}_0=\hbar v_{\rm F}\overrightarrow{\sigma}\cdot{\bf k}+mv_{\rm F}^2\sigma_3$. There are two eigenvalues $\pm E_{\bf k}$ where $E_{\bf k}=\sqrt{\varepsilon_{\bf
k}^2+\Delta^2}$ is the spectrum of particle with $\Delta=mv^2_{\rm F}$ is the gap energy. The results of gapless graphene can be obtained by setting $\Delta=0$.
It should be noted that the in-plane magnetic field couples not only to the spin degrees of freedom in a quasi 2DEG which leads to the spin polarization of the carriers, but also to the orbital motion of
the carriers~\cite{sarma,sarma1,tanatar,senator,experimental} due to the finite quantum well thickness.
However, because of the absence of thickness in graphene, the applied
$B$ couples only with spin of the carriers and leads to the spin polarization of them.
Due to the Zeeman spin-splitting effect we have a shift in the Fermi
wave vector for up and down spins
$k_{{\rm F}\sigma}=k_{\rm F}(1+\sigma\zeta)^{1/2}$ where
$k_{\rm F}=\sqrt{4\pi n/g_sg_v}$ is the unpolarized Fermi wave vector, $n$ is the 2D electron density and
$g_s=2$ and $g_v=2$ are the spin and valley degeneracy, respectively.
The coupling constant in graphene sheets is density independent and
given by $\alpha_{gr}=g_sg_ve^2/\epsilon\hbar v_F$ where $\epsilon$
is the average dielectric constant of the substrate and air.
For a typical substrate (e.g. {\rm SiC} or {\rm SiO$_2$}) the dielectric constant is between 1 and 2.

The total energy per particle in the presence of an in-plane
magnetic field $B$ as a function of the density $n$, the spin
polarization $\zeta$, the gap energy $\Delta$ and the coupling
constant $\alpha_{gr}$ takes the following form
\begin{eqnarray}\label{energy}
\varepsilon_{tot}(n,\zeta,\Delta,B)&=&\varepsilon_{kin}(n,\zeta,\Delta)+\varepsilon_x(n,\zeta,\Delta)\nonumber\\
&+&\varepsilon_c(n,\zeta,\Delta)+\varepsilon_Z(\zeta,B).
\end{eqnarray}
where
\begin{eqnarray}
\varepsilon_{kin}(n,\zeta,\Delta)&=&\frac{g_v}{6\pi n\hbar^2v_{\rm F}^2}\{[\hbar^2v_{\rm F}^2k_{\rm F}^2(1+\zeta)+\Delta^2]^{3/2}\nonumber\\
&+&[\hbar^2v_{\rm F}^2k_{\rm F}^2(1-\zeta)+\Delta^2]^{3/2}
-2\Delta^3\}
\end{eqnarray}
is the kinetic energy per particle and
\begin{eqnarray}
&&\varepsilon_x(n,\zeta,\Delta)=-\frac{1}{4\pi n}\int\frac{d^2{\bf
q}}{(2\pi)^2}V_q\int_0^\infty d\omega\,\,\, \nonumber\\
&[&{\chi_\uparrow^{(0)}({\bf
q},i\omega,\zeta,\Delta)+\chi_\downarrow^{(0)}({\bf
q},i\omega,\zeta,\Delta)}],
\end{eqnarray}
is the exchange energy.  $\chi^{(0)}({\bf
q},i\omega,\zeta,\Delta)$ is the zero temperature noninteracting polarization
function for doped graphene \cite{qaium} which is given by
\begin{widetext}
\begin{eqnarray}\label{eq:final_result}
\chi^{(0)}_{\sigma}({\bf q},i \omega,\zeta,\Delta)&=&-\frac{g_v}{2\pi
\hbar^2v^2}\{\mu_{\sigma}- \Delta+\frac{
\varepsilon_q^2}{2}\left[\frac{\Delta}{{\varepsilon_q^2+\hbar^2 \omega^2}}+\frac{1}{2\sqrt{\varepsilon_q^2+\hbar^2\omega^2}}
(1-\frac{4\Delta^2}{\varepsilon_q^2+\hbar^2\omega^2})\tan^{-1}(\frac{\sqrt{\varepsilon_q^2+\hbar^2\omega^2}}{2\Delta})\right]\nonumber\\&-&
\frac{\varepsilon_q^2}{4\sqrt{\hbar^2\omega^2+\varepsilon_q^2}}\Re
e\left [(1-\frac{4\Delta^2}{\varepsilon_q^2+\hbar^2\omega^2})\{\sin^{-1}(\frac{2\mu_{\sigma}+i
\hbar\omega}{\varepsilon_q \sqrt{1+\frac{4\Delta^2}{\varepsilon_q^2+\hbar^2\omega^2}}})-\sin^{-1}(\frac{2 \Delta+i
\hbar\omega}{\varepsilon_q \sqrt{1+\frac{4\Delta^2}{\varepsilon_q^2+\hbar^2\omega^2}}})\}\right]\nonumber\\&-&
\frac{\varepsilon_q^2}{4\sqrt{\hbar^2\omega^2+\varepsilon_q^2}}\Re
e\left [(\frac{2\mu_{\sigma}+i\hbar\omega}{\varepsilon_q})\sqrt{(1+\frac{4\Delta^2}{\varepsilon_q^2+\hbar^2\omega^2})-(\frac{2\mu_{\sigma}+i\hbar\omega}
{\varepsilon_q})^2}\right ]\nonumber\\&+&
\frac{\varepsilon_q^2}{4\sqrt{\hbar\omega^2+\varepsilon_q^2}}\Re
e\left [(\frac{2\Delta+i\hbar\omega}{\varepsilon_q})\sqrt{(1+\frac{4\Delta^2}{\varepsilon_q^2+\hbar^2\omega^2})-(\frac{2\Delta+i\hbar\omega}
{\varepsilon_q})^2}\right ]\}~,
\end{eqnarray}
\end{widetext}

where $\mu_{\sigma}=\sqrt{\hbar^2v^2{k_{{\rm F}\sigma}}^2+\Delta^2}$ and $V_q=2\pi
e^2/\epsilon q$ is the 2D Coulomb interaction. Moreover, the correlation
energy per particle~\cite{asgari} in RPA is given by
\begin{eqnarray}
&&\varepsilon_c(n,\zeta,\Delta)=-\varepsilon_x(n,\zeta,\Delta)+\frac{1}{2\pi n}\int\frac{d^2{\bf
q}}{(2\pi)^2}\int_0^\infty d\omega\nonumber\\
\ln&[&1-V_q(\frac{\chi_\uparrow^{(0)}({\bf
q},i\omega,\zeta,\Delta)+\chi_\downarrow^{(0)}({\bf
q},i\omega,\zeta,\Delta)}{2})],
\end{eqnarray}
and finally the Zeeman energy per particle is $\varepsilon_Z(\zeta,B)=-\mu_B\zeta B$
where $\mu_B$ is the Bohr
magneton. In the above equation we have used the fluctuation-dissipation
theorem~\cite{Giuliani_and_Vignale}. In order to make the exchange and correlation energies finite, we might subtract~\cite{qaium,asgari} the vacuum polarization
energy contributions from the exchange and correlation energies $\delta
\varepsilon_{x(c)}(k_{\rm F}\neq0)=\varepsilon_{x(c)}(k_{\rm F})-\varepsilon_{x(c)}(k_{\rm F}=0)$. Due to the conserved number of states in the Brillouin zone, we do need an
ultraviolet momentum cut-off $k_c$ which is approximated by
$\pi k_c^2=(2\pi)^2/\mathcal{A}_0$ where
$\mathcal{A}_0$ is the area of the unit cell in the honeycomb lattice. The dimensionless parameter $\Lambda$ is defined as $k_c/k_{\rm F}$.

The total energy per
particle for a gapless graphene in the noninteracting electron scheme is given by
\begin{equation}
\varepsilon^0_{tot}(n,\zeta,B)=\frac{g_v\varepsilon_{\rm F}k_{\rm F}^2}{6\pi n}[(1+\zeta)^{3/2}+(1-\zeta)^{3/2}]-\mu_B\zeta B
\end{equation}
where
$\varepsilon_{\rm F}=\hbar v_{\rm F}k_{\rm F}$ is the Fermi energy of the gapless
graphene. The minimum of the noninteracting energy as a function of spin
polarization occurs at
$\zeta^*_0=2\mu_B
B(\varepsilon_{\rm F}^2-\mu_B^2B^2)^{1/2}/\varepsilon_{\rm F}^2$.
Setting $\zeta^*=1$ allows us to determine the critical magnetic field $B_{c0}(n)$ corresponding to the fully spin polarize the system. The critical magnetic field for the noninteracting system is
$B_{c0}={\varepsilon_{\rm F}}/{\sqrt{2}\mu_B}$.

To calculate $\zeta^*(B)$ for the interacting case, the total energy in Eq.~(\ref{energy}) needs to be minimized with
respect to $\zeta$ and then the critical magnetic
field $B_c$ can be found for the fully spin polarized case. At a finite magnetic field the energy minimum occurs at nonzero polarization $0 < \zeta^*< 1$. Beyond the critical field the energy minimum is at $\zeta^*=1$ and the system is fully polarized. In general, the critical magnetic field takes the form \begin{eqnarray}\label{bc}
\frac{B_c}{B_{c0}}
=\frac{\sqrt{2}}{2 \varepsilon_F}\left\{[(2\varepsilon^2_F+\Delta^2)^{1/2}-\Delta]
+2\frac{\partial\delta\varepsilon_{xc}}{\partial
\zeta}|_{\zeta=1}\right\}
\end{eqnarray}

\section{Numerical Results}

We now turn to the presentation of our numerical results. We consider
$\alpha_{gr}=1$ which is a appropriate value
for graphene placed on the {\rm SiC} substrates and
we choose the gap values between 0 and 100 meV which is observed in typical experiments.
\begin{figure}
\includegraphics[width=7cm]{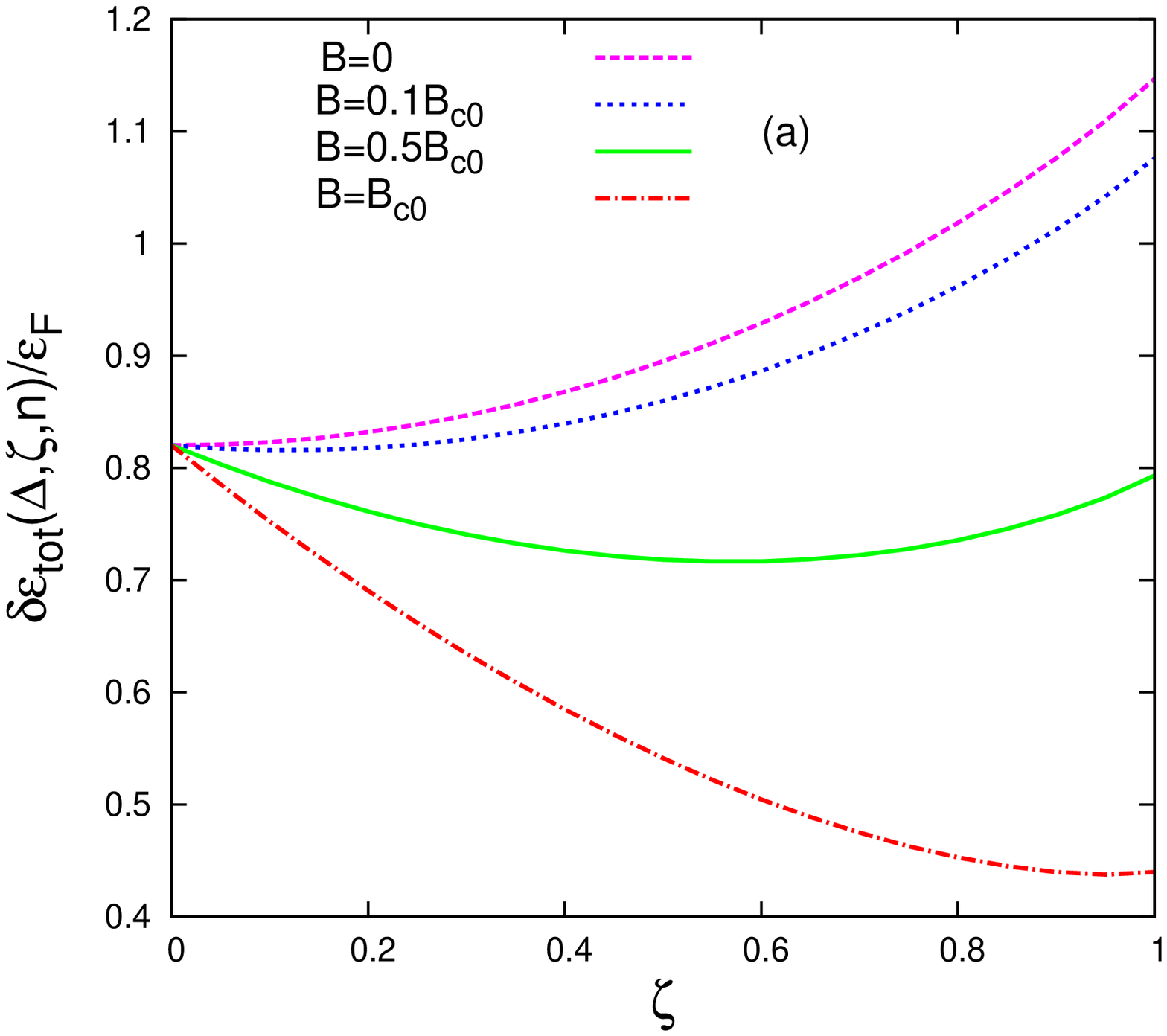}
\hfill
\includegraphics[width=7cm]{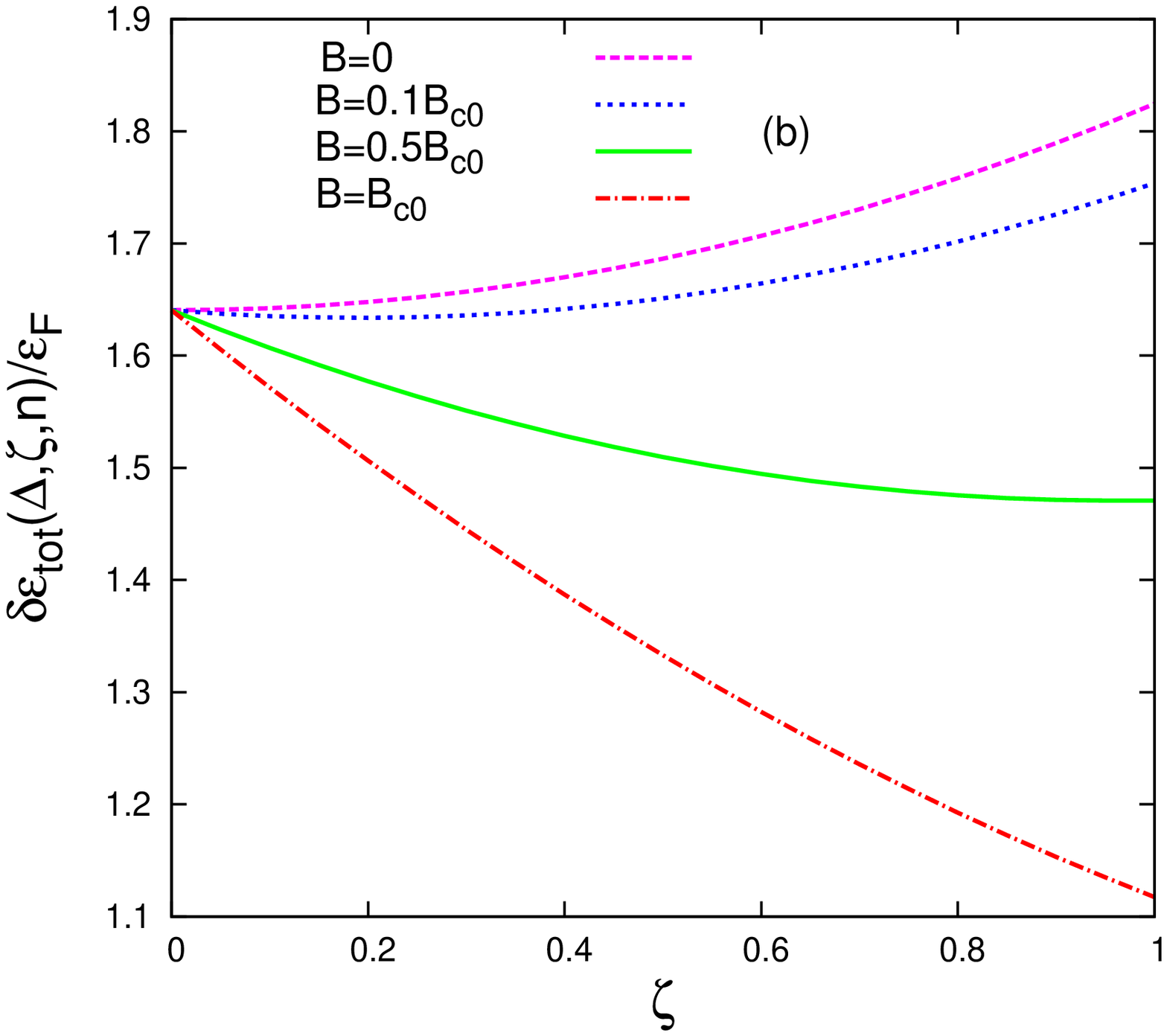}
\caption{(color online). Total energy as a function of spin
polarization for various magnetic fields for (a): $\Delta=0$ and (b): $\Delta=100$meV at $\Lambda=100$. }
\end{figure}

In Fig.~1 we plot the exchange ( in the inset: correlation ) energy of graphene as a function of $\zeta$ for
a range of $\Delta$ values.  The exchange energy
is positive because our regularization procedure implicitly selects the chemical potential
of undoped graphene as the zero of energy. It would be noted that $\partial \delta\varepsilon_{x}/\partial\zeta |_{\zeta=1}>0$ at small $\Delta$ and it changes the sign at large gap values. The slope of exchange and correlation energies with respect to $\zeta$ around $\zeta=1$ have opposite signs.
Note that both $\delta \varepsilon_{\rm x}$ and
$|\delta \varepsilon_{\rm c}|$ have the same density dependence and they both increase with decreasing  density.~\cite{asgari} These arguments will be useful in describing the the critical magnetic field given by Eq.~\ref{bc}.

In Fig.~2, we plot the calculated critical field $B_c$ which polarizes the
quasiparticles for the interacting case in the units of the critical field for noninteracting Dirac
massless fermions $B_{c0}$, as a function of inverse square root of
the density. The critical field increases by decreasing
the density of carriers for massless case due to the impact of exchange energy ( see Eq.~\ref{bc}). This particular feature
is in contrast to the 2DEG in which the reduction of carrier density
leads to the decline of the critical field.\cite{sarma,tanatar} This
distinguished behavior is a direct result of the chirality in the
massless Dirac fermions and interband interaction features. The critical fields, on the other hand, have a non monotonic behavior at small gap values. It increases
by decreasing the density till it reaches to the maximum value and
then decreases at very low density values because of the competition between the exchange and correlation energy contributions, as shown in Fig.~1. We have found that $B_c$ vanishes at about $n\sim 10^8$ cm$^{-2}$ for $\Delta=100$ meV. This is a
similar behavior to the conventional 2DEG where system goes to the fully polarized state spontaneously at a special density, $r_s\approx25.5$ calculated within Monte Carlo simulations~\cite{attaccalite} or $r_s\approx5.5$ base on RPA calculation.~\cite{sarma}

Moreover, at low density region associated with large $\Lambda$, we found no indication for a spontaneous magnetic phase transition for small $\Delta$ even at large coupling constants. These results are in contrast to the results reported in Ref.~[\onlinecite{peres}] where the exchange term, the Hartree-Fock theory was only used. In that work, the authors found that exchange interactions between Dirac fermions can stabilize a ferromagnetic phase at low doping when the coupling is sufficiently large.
We have not found any evidence for this instability using RPA calculations. The RPA is a minimal dielectric scheme that allows quantitative predictions beyond the Hartree-Fock theory. In the present case of a two-dimensional electron gas on a graphene sheet, the Hartree-Fock exchange contribution to the ground-state energy is positive. In our work we clearly show that the RPA correlation energy is negative.

Furthermore, it is shown~\cite{dharma-wardana} that the kinetic energy enhancement of the spin-polarization phase nearly cancels the exchange enhancement and the correlation energy plays a dominant residual role. Therefore, the inclusion of the correlation energy suppresses the spin-polarized phase found in the exchange only calculation in gapless graphene. In examining the tendency of the system to develop magnetic order in the presence of electron-electron interactions it is thus crucially important to include both exchange and correlation contributions.

In Fig.~3 we plot the calculated total ground state energy, $\delta
\varepsilon_{tot}(k_{\rm F})=\varepsilon_{tot}(k_{\rm F})-\varepsilon_{tot}(k_{\rm F}=0)$ in units of the
Fermi energy for massless Dirac fermions $\varepsilon_{\rm F}$ as a
function of the spin polarization parameter $\zeta$. The results are shown for various
magnetic fields at $(a) \Delta=0$ and $(b) \Delta=100$meV. In both cases
the minimum energy occurs at paramagnetic state, namely $\zeta^*=0$ in the absence of magnetic field but as
$B$ increases the minimum energy shifts to
non-zero spin polarization and $\zeta^*$ increases continuously
to the ferromagnetic phase ($\zeta^*=1$) at $B=B_c$. For $B>B_c$
the system remains in the ferromagnetic phase. This indicates a continuous-phase
transition (Stoner type) from para- to- ferromagnetic phase in
the presence of magnetic field for each density value whereas a first-order
phase transition for whole of the density range is predicted for 2DEG.~\cite{sarma}

\begin{figure}
\includegraphics[width=7.5cm]{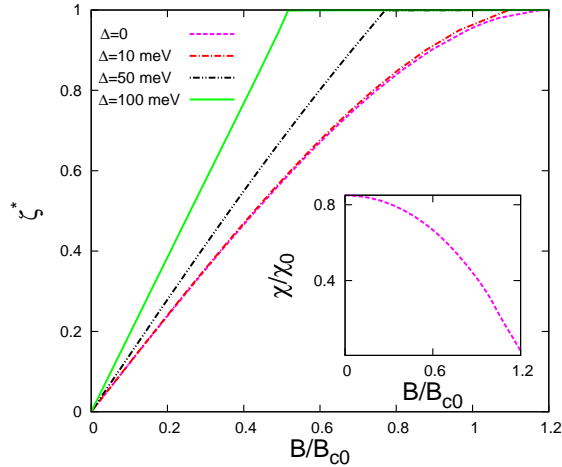}
\caption{(color online). Spin polarization as a function of the magnetic field
for several energy gap values at $\Lambda=100$. The inset: The spin susceptibility as a function of the magnetic
field for $\Delta=0$. } \label{}
\end{figure}

In Fig.~4 we plot the magnetization $\zeta^*(B)$ as a function
of the applied magnetic field $B$.~\cite{sarma,tanatar} Clearly there is
no longer jump in the magnetization at $B=B_c$ due to
a continuous-phase transition in graphene. In the conventional 2DEG, the transition to the ferromagnetic state near the critical magnetic field value happens with a discrete jump in the polarization indicating a first order transition to the fully polarized state. The magnetization
$\zeta^*$ is a semi-linear function of the magnetic field versus the large gap values.

A quantity of interest which can be accessed
experimentally is the non-linear spin susceptibility of the system defined as $\chi/\chi_0(B=0) =\frac{\varepsilon_{\rm F}}{2\mu_B}\partial\zeta^*/\partial B$ where $\chi_0$ is the Pauli susceptibility. The spin susceptibility decreases nonlinearly by increasing the magnetic field at small $\Delta$ values, showing that the polarizability of the system decreases. This feature should be verified by magneto-resistance measurements through the polarization field $B_c$.

Another important thermodynamic quantity is the compressibility, $\kappa$ which
yields interesting features when graphene is subjected to an in-plane magnetic field.
The exchange energy is positive while the correlation energy is negative. This has important implications on the thermodynamic properties. The compressibility can be calculated from its definition, $\kappa^{-1}=n^2\partial^2(n\delta\varepsilon_{tot})/\partial n^2$.
In Fig.~5 we have shown the inverse of compressibility of gapless
graphene as a function of the inverse square root of density for unpolarized and fully polarized
states. $\kappa_0/\kappa$ increases with decreasing density
at small gap energy. This behavior is in contrast to the conventional 2DEG. The compressibility of noninteracting gapless graphene is
$\kappa_0=2/(n\varepsilon_{\rm F})$. The exchange energy tends to reduce the compressibility while correlations tend to enhance it. We found that
at the given $B_c(\Lambda=5)$ which is associated with a special density, $\kappa_0/\kappa$ starts from $\zeta=1$
and slowly tends to the paramagnetic results of the gapless one. This special behavior is a consequence of the fact that $B_c(\Lambda<5)$ is smaller than $B_c(\Lambda=5)$ as shown in Fig.~2. However at gapped graphene, say $\Delta=100$, we chose a value of the critical magnetic field $B_c(\Lambda=400)$ and  observed that $\kappa_0/\kappa$ switches to its fully polarized system value with a kink-like behavior as shown in the Fig.~6. This feature is a consequence of the fact that $B_c(\Lambda<400)$ is larger than $B_c(\Lambda=400)$. This suggests that in the compressibility measurements the effect of the polarizing magnetic filed could be discerned. The physical reason for having two different behaviors at small and large $\Lambda$ is that the critical magnetic field behaves in different ways
at small and large energy gap values. Note that at very large gap energy, $\kappa_0/\kappa$ decreases by increasing $n^{-1/2}$. The non-monotonic behaviors of $\kappa_0/\kappa$ with respect to $\Delta$ is due to the comparison between the exchange energy and the correlation energy as a function of gap values.
\begin{figure}
\includegraphics[width=7.5cm]{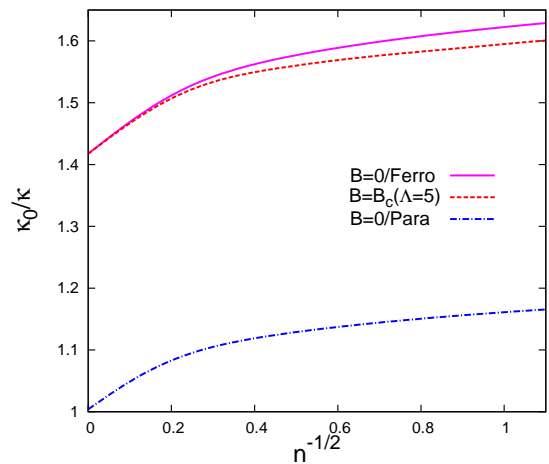}
\caption{(color online). Compressibility of gapless graphene as a
function of inverse square root of density (in units of 10$^{-6}$ cm) for both fully spin polarized and unpolarized states.} \label{}
\end{figure}

\section{Conclusion}

In summary, we study the effects of in-plane magnetic field on
the ground state properties of both gapless and gapped graphene where the conduction band is partially occupied. The present work is demonstrative the increasing behavior of the critical field in which the system becomes fully spin
polarized by decreasing the density for gapless
graphene. Accordingly, it means that there is no longer a spontaneous para-to- ferromagnetic phase transition for gapless graphene at zero-magnetic field. The critical magnetic
field decreases by decreasing the density at large gap values. Quite interestingly, we find a continuous
quantum magnetic phase transition for the
whole range of the density at zero temperature. The novel in-plane magnetic field dependence of charge and spin susceptibilities are obtained. The inverse compressibility as a function of inverse density exhibits a crossover from the fully polarized state to the paramagnetic case for gapless graphene, which should be identifiable experimentally.

\begin{figure}
\includegraphics[width=7.5cm]{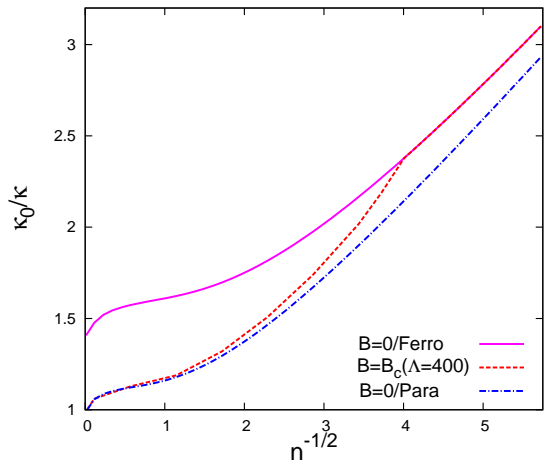}
\caption{(color online). Compressibility of gapped graphene with $\Delta=100$meV as a
function of inverse square root of density (in units of 10$^{-6}$ cm) for both fully spin polarized and unpolarized states.} \label{}
\end{figure}

It is worth noticing that the aforementioned results are in contrast
to those results calculated for a conventional 2DEG
due to the effect of interband interaction. We suggest these results should be verified by experiments.

It is recently shown that ripples in graphene induced a gauge
field. It is convenient to emphasize that the study of the effects of parallel
magnetic field on physical quantities in graphene sheets at the presence of such an
induced gauge filed is an interesting problem which might be taken into account.

\section{Acknowledgment}
We would like to thank F. K. Joibari for useful comments. A. Q is supported by IPM grant.

\end{document}